# Ab initio Investigations on the Electronic Properties and Stability of Cu-Substituted Lead Apatite (LK-99) Family with Different Doping Concentrations (x=0, 1, 2)

Songge Yang[1], Guangchen Liu[1], Yu Zhong[1]

[1]Mechanical and Materials Engineering, Worcester Polytechnic Institute, Worcester, MA 01609

(Date: August 28, 2023)

*Abstract*

*The pursuit of room-temperature ambient-pressure superconductivity in novel materials has sparked interest, with recent reports suggesting such properties in Cu-substituted lead apatite, known as LK-99. However, these claims lack comprehensive experimental and theoretical support. In this study, we address this gap by conducting ab initio calculations to explore the impact of varying doping concentrations (x = 0, 1, 2) on the stability and electronic properties of five compounds in the LK-99 family. Our investigations confirm the isolated flat bands that intersect the Fermi level in LK-99 ($Pb_9Cu(PO_4)_6O$:Cu$<Pb(1)>$). In contrast, the other four parent compounds exhibit insulating behavior with wide band gaps. X-ray diffraction spectra based on the DFT simulations at 0K confirm the presence of Cu substitution on Pb(1) sites in the originally synthesized LK-99 sample, while an extra peak suggests potential alternative like $Pb_8Cu_2(PO_4)_6$ phases due to compositional variations in the original LK-99 samples. Furthermore, the LK-99 structure undergoes substantial lattice constriction, resulting in a significant 5.5% reduction in volume and 6.8% in area of two mutually inverted triangles formed by Pb(2) atoms. Meanwhile, energy calculations reveal a marginal energy preference for substituting Cu on Pb(2) sites over Pb(1) sites, with a difference of approximately 0.010 eV per atom (roughly 0.9645 k/mol). Intriguingly, at pressures exceeding 73 GPa, stability shifts towards LK-99 containing Cu substitutions on Pb(1) sites. Despite exhibiting higher electronic conductivity than parent compounds, $Pb_9Cu(PO_4)_6O$:Cu$<Pb(1)>$ falls short of the conductivity levels observed in metals or advanced oxide conductors with the simulation based on the Boltzmann transport theory.*

## 1. Introduction

A superconductor with a high critical temperature ($T_C$) [1] has attracted extensive attention in the field of condensed matter physics due to its superconductivity [2], complete diamagnetism (Meissner effect) [3, 4], magnetic flux quantization [5, 6], zero electronic resistance below $T_C$ [7], and etc. The first class of superconductors that were considered to be high-$T_C$ were the cuprates, which were discovered by Bednorz and Müller in 1987 [8]. Subsequently, the cuprates have been succeeded by several novel categories, such as the Fe-pnictides introduced in 2008 [9] and the nickelates [10]. While significant efforts, i.e., *MgB$_2$* ($T_C$=39K, ambient pressure) [11], *La$_3$Ni$_2$O$_7$* ($T_C$=80K, 14.0–43.5 GPa) [12], *Hg-Ba-Ca-Cu-O system* ($T_C$=133K, ambient pressure) [13], *LaH$_{10}$* ($T_C$=250K, 170Gpa) [14], etc., have been made in the discovery of high-$T_C$ superconductors, a clear path toward achieving room-temperature superconductivity under ambient pressure has remained challenging.

A groundbreaking discovery of possible room-temperature superconductivity was recently reported by Lee et al. [15, 16], claiming their significant milestone achievement by creating the first room-temperature ambient-pressure superconductor using Cu-doped lead-apatite (LK-99). The author argued that the superconductivity of LK-99 originates from the structural distortion by volume shrinkage caused by the Cu substitution on Pb(1) sites (Wyckoff position 4f). Later on, Griffin et al. [17] performed density functional theory calculations on the LK-99 materials, identifying the correlated isolated flat bands at the Fermi level, a characteristic often associated with high transition temperatures in the established superconducting family. This finding provides empirical support for the intrinsic potential of the LK-99 material to exhibit superconducting properties from a physical perspective. Unfortunately, recent attempts to find superconductivity in LK-99 have been unsuccessful [18, 19], thus tempering the initial optimism. Some researchers [20, 21] have also raised questions about the superconductivity of LK-99. Zhu et al. [21] claimed that LK-99 is not a single phase, and the superconducting-like behavior most likely originates from a magnitude reduction in resistivity caused by the first-order transition of Cu$_2$S. Nevertheless, this particular material system possesses inherent intrinsic interest, affording abundant opportunities for theoretical investigation.

So far, there are many opening questions regarding how the doping of Cu will affect the stability and electronic conductivity of lead apatite materials. In the present work, we set out to investigate the effect of doping concentrations (x = 0, 1, 2) on the stability and electronic properties of LK-99 family materials through *ab initio* calculations. Besides confirming the crystal structure of LK-99, we also included other four family compounds with the consideration of different doping concentration and substitution sites, which might form during the synthesis process. In particular, a lattice distortion analysis was conducted to examine whether the substitution of Pb with Cu would contribute to volume shrinkage and induce local distortions. Our comprehensive analyses pave the path for a deeper comprehension of the isolated flat bands observed within LK-99. By addressing these questions, we endeavor to enrich our insights into the realm of superconductors, thereby advancing our understanding of how Cu doping influences stability and electronic properties in lead apatite materials.

## 2. Computational details

The *ab initio* calculations were applied using the Vienna ab initio simulation package (VASP) [22, 23]. Throughout the study, we have utilized the generalized gradient approximation (GGA) [24] with the Perdew-Burke-Ernzerhof (PBE) exchange-correlation function [25] using the projector augmented wave method. The k-point meshes with a density of no less than 5000 pra (per-reciprocal atom) were used to sample the Brillouin zones. The accurate total energy calculations were applied using the linear tetrahedron method with Blöchl's corrections [26]. The total energies obtained by relaxation calculations were converged to $10^{-6}$ eV/cell with a 1.75 times plane-wave cutoff energy suggested by the corresponding element pseudopotentials [27-31]. Meanwhile, the static calculations, which converged to $10^{-7}$ eV/cell with 520 eV cutoff energy, were applied after each relaxation procedure. Finally, the calculated static energies were fitted by the Birch-Murnaghan equation of state [32] for each structure to get a more accurate total energy. Considering the possible magnetic effect of the Cu atom, the spin polarization was switched on for the calculation of Cu-doped LK-99. Although the *ab initio* calculations provide good agreement with the experiments for metallic systems as well as a reasonable trend for the geometry and reactivity of metal oxides, this method fails to obtain the correct electronic structure for the strongly correlated system due to errors associated with on-site Coulomb and exchange interactions [33, 34]. These errors limit the applicability of the *ab initio* calculations in late-transition metal oxides [35]. A possible workaround is using the GGA+U method, which considers local corrections within selected orbitals. In the present work, the Hubbard-U corrections were applied to account for the underlocalization of Cu-*d* states. The value U=4 eV [17], which gives the lattice parameters within 1% of the experiment [15, 17, 36], is selected for Cu. Throughout this study, the graphical visualization package VESTA [37] was used to check the lattice distortion for the doped structures.

## 3. Results and Discussions

*3.1. Equation of state calculations of $Pb_{10-x}Cu_x(PO_4)_6O$ (x=0,1) and $Pb_8Cu_2(PO_4)_6$ (x=2)*

In order to get the accurate equilibrium volume, the total energy of $Pb_{10-x}Cu_x(PO_4)_6O$ (x=0,1) and $Pb_8Cu_2(PO_4)_6$ (x=2) were calculated and fitted based on the 4-parameter Murnaghan EOSs [32] as shown in Eq.(1):

$$E^\sigma(V) = a + \frac{B_0 V}{B_0'}\left(1 + \frac{\left(\frac{V_0}{V}\right)^{B_0'}}{B_0' - 1}\right) \quad (1)$$

In Eq.(1), $a = E_0 - \frac{B_0 V_0}{B_0' - 1}$, the parameters $V_0$, $E_0$, $B_0$, and $B_0'$ represent the equilibrium volume, energy, bulk modulus, and its first derivation with respect to pressure, respectively. The fitted results are shown in *Figure 1*, and the equilibrium volumes ($V_0$) are shown in Table 1, including the comparison with the other peoples' work [5, 7, 15, 38].

In this section, we investigated five cases: $Pb_{10}(PO_4)_6O$, $Pb_9Cu(PO_4)_6O$:Cu<Pb(1)> (with one copper atom substituted at the Pb(1) site), $Pb_9Cu(PO_4)_6O$:Cu<Pb(2)> (with one copper atom substituted at the Pb(2) site), $Pb_8Cu_2(PO_4)_6$:2Cu<2Pb(1)> (with two copper atoms substituted, both at Pb(1) sites), and $Pb_8Cu_2(PO_4)_6$:2Cu<1Pb(1), 1Pb(2)> (with two copper atoms substituted, one at Pb(1) site and one at Pb(2) site). The detailed explanations of two different Pb sites were discussed in *Section 3.2*.

Based on the result of the equation of state calculations, it has been determined that, as a general trend, the equilibrium volume of lead apatite tends to decrease with the doping of Cu. Particularly noteworthy is the effect of substituting Cu on the Pb(1) site, which leads to a more significant lattice contraction compared to Cu substitution on the Pb(2) site. Among all these cases, the $Pb_9Cu(PO_4)_6O$:Cu<Pb(1)> has the largest lattice contraction, showing a 5.5% reduction in volume. This volume shrinkage is much higher than the value reported by Lee et al. [15] (0.48%), which could be attributed to the overestimation of the volume of $Pb_{10}(PO_4)_6O$ induced by PBE function. Regarding the calculated total energy, it is found that the Cu substitution on Pb(2) sites is more energetically favorable than Pb(1) sites by around 0.010 eV per atom (0.9645 kJ/mol). This finding is consistent with the results reported by Griffin et al. [17], Shen et al.[39], and Lai et al.[5].

Table 1 exhibits the fitted equilibrium volume combined with the comparison of other people's work [5, 7, 15, 38]. The investigation reveals that, when compared with experimental data, the *ab initio* calculations from both our study and investigations from other groups [5, 7] yield a larger volume estimation of $Pb_{10}(PO_4)_6O$. In the case of $Pb_9Cu(PO_4)_6O$, it is observed that, after the consideration of Hubbard-U correction, our predicted volume of $Pb_9Cu(PO_4)_6O$<Pb(1)> shows excellent agreement with the result of LK-99 [15]. It has been speculated that [15] the replacement of Cu atom within LK-99 at the Pb(1) site. However, according to the total energy calculations from the present work, Cu substitution on Pb(2) sites is thermodynamically more favorable compared to substitution on Pb(1) sites. In order to investigate the preferential occurrence of Cu substitution either at the Pb(1) or Pb(2) sites, the powder diffraction pattern analysis will be discussed in the next section, in direct comparison with experimental X-ray diffraction (XRD) data.

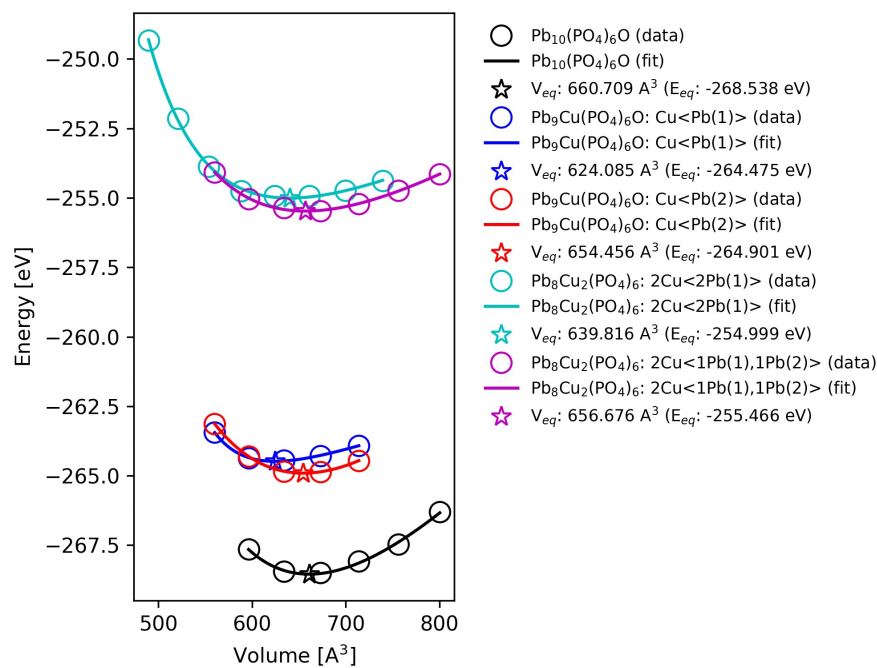

Figure 1 The energy-volume curve of $Pb_{10-x}Cu_x(PO_4)_6O$ (x=0,1) and $Pb_8Cu_2(PO_4)_6$ (x=2)

Table 1 The fitted equilibrium volume ($A^3$) including the comparison of other people's work

| Structures | This work (DFT) | Lai et al. [5] (DFT) | Hao et al.[7] (DFT) | Exp [15, 38] |
|---|---|---|---|---|
| $Pb_{10}(PO_4)_6O$ | 660.709 | 651.02 | 642.75 | 626.25 |
| $Pb_9Cu(PO_4)_6O$:Cu<Pb(1)> | 624.085 | 632.94 | 610.75 | 623.24 (LK-99) |
| $Pb_9Cu(PO_4)_6O$:Cu<Pb(2)> | 654.456 | | | |

| | |
|---|---|
| $Pb_8Cu_2(PO_4)_6$:2Cu<2Pb(1)> | 639.816 |
| $Pb_8Cu_2(PO_4)_6$:2Cu<1Pb(1), 1Pb(2)> | 656.676 |

*3.2. XRD analysis*

The XRD spectra from the 2θ region from 18° to 34° for all the five different compounds were included in Figure 2 for comparison. They are based on the structure relaxed at 0K with the DFT simulations, which should have a slight peak shift in comparison with the regular experimental XRD peaks taken at room temperature. The calculations of diffraction pattern were performed by the VESTA software. The target material is selected as CuKα (λ=1.5406A). The XRD spectra from the pure $Pb_{10}(PO_4)_6O$ is adopted as the benchmark and the individual peaks were labeled as a, b, c, d, e, f, g, h, and i. It shows that the pure $Pb_{10}(PO_4)_6O$ has the highest symmetry with the main peak f has the highest intensity. By adding one Cu atom with $Pb_9Cu(PO_4)_6O$:Cu<Pb(1)>, there is no apparent symmetry change, however, all the peaks shift to the right, which is an indication of the lattice parameter shrinkage by adding the Cu atom. Especially, the main peak f at 30.19° from the pure $Pb_{10}(PO_4)_6O$ is shifted to 30.64°.

Meanwhile the stable x=1 structure by adding one Cu atom to the Pb(2) site greatly destroyed the symmetry; in which the peak broadening should be observed from experiments. Similarly, it is observed that by adding 2 Cu atoms to the structure, the XRD spectra can be very different. Significant peak broadening should be observed from the $Pb_8Cu_2(PO_4)_6$:2Cu<1Pb(1), 1Pb(2)> structure. In comparison with the experimental XRD spectra from the original LK-99 sample [15, 16], it can be concluded that LK-99 should have the $Pb_9Cu(PO_4)_6O$:Cu<Pb(1)> structure. Moreover, additional peaks observed with angle slightly lower than d peak and one between peaks d and e [15, 16]. It indicates that the LK-99 samples are not pure and should have phases/structures other than the well-known $Cu_2S$ impurity phase. The additional peak between d and e may due to the existence of $Pb_8Cu_2(PO_4)_6$:2Cu<2Pb(1)> as the LK-99 Pb:Cu ratio is not exactly 9:1 and it is possible more than one Cu atom enters the $Pb_{10}(PO_4)_6O$ structure.

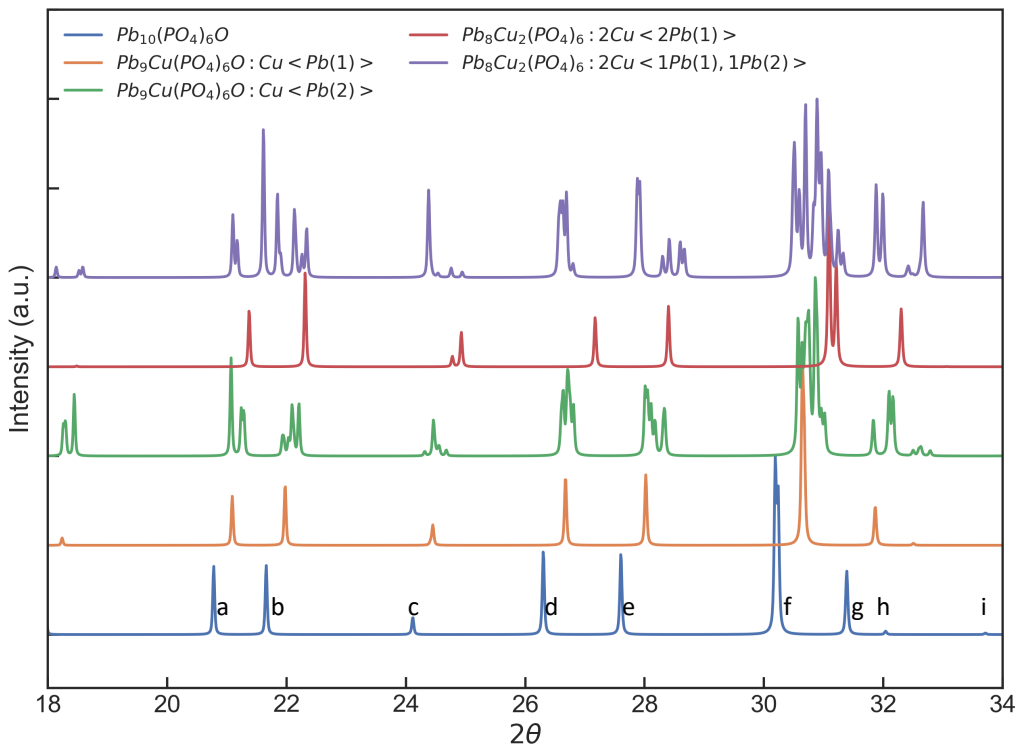

*Figure 2 XRD spectra for the 5 compounds in the LK-99 family investigated in the present work.*

3.3. Electronic properties of $Pb_{10-x}Cu_x(PO_4)_6O$ (x=0,1) and $Pb_8Cu_2(PO_4)_6$ (x=2)

3.3.1 $Pb_{10}(PO_4)_6O$

The fully relaxed crystal structures of pure $Pb_{10}(PO_4)_6O$ are shown in Figure 3(a). Within the crystal lattice of lead apatite, there are two non-equivalent lead (Pb) sites, denoted as Pb(1) and Pb(2), exhibiting distinct symmetry configurations. The Pb(1) atoms (Wyckoff position 4f) exhibit a hexagonal arrangement (green hexagon), while the Pb(2) atoms (Wyckoff position 6h) assume the formation of two mutually inverted triangles (yellow triangles). Previous studies have reported [17] that within the structure of lead apatite, Pb(1) establishes the overall framework, while Pb(2) plays a crucial role in Pb-O connectivity and polyhedral tilting, with the chiral arrangement of its lone pairs leading to asymmetric oxygen coordination and the propagation of structural distortions. The calculated band structures and electronic density of state (DOS) for each element are shown in Figure 3(b) and (c) and *Figure 4*, respectively. It can be found that $Pb_{10}(PO_4)_6O$ is a nonmagnetic insulator with a 2.57 eV band gap (*Figure 4*). The DOS reveals that the valence bands with the energy range near the Fermi level originate primarily from the dominant O-2p state, supplemented by minor Pb-(6s + 6p) states [5]. Specifically, two flat bands, marked in red as shown in Figure 3(c), are observed near the Fermi energy. Flat bands are often considered advantageous for strongly correlated phases due to the prevalent influence of interaction strength over single-particle bandwidth. Jiang et al. [40] proposed that if the flat band reaches the limit (t → 0), the system transforms into a fully decoupled lattice (the atomic limit), exhibiting complete paramagnetism. However, when the interaction strength exceeds that of isolated bands yet remains below the gap between these bands and their complements, a distinct array of phases can emerge. In the case of repulsive interactions, ferromagnetism can be rigorously established [41, 42], while attractive interactions can lead to superconductivity (or phase separation) [42-44].

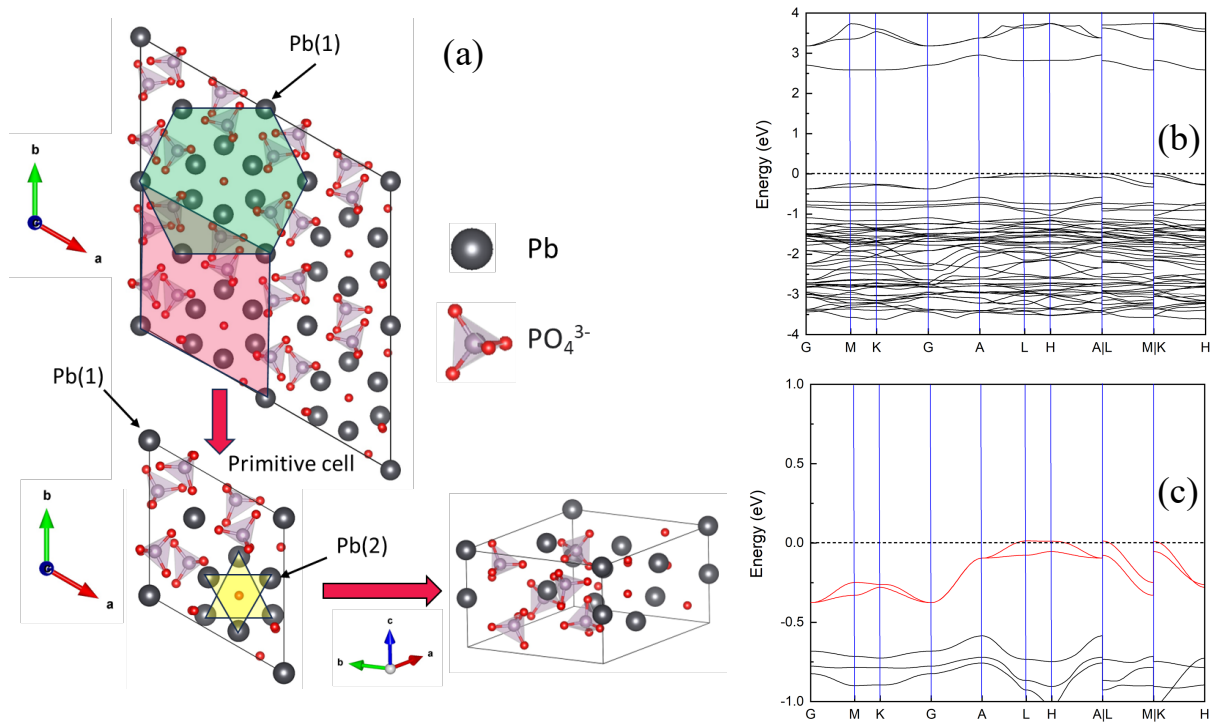

Figure 3 The structure configurations and band structure of the $Pb_{10}(PO_4)_6O$ compound

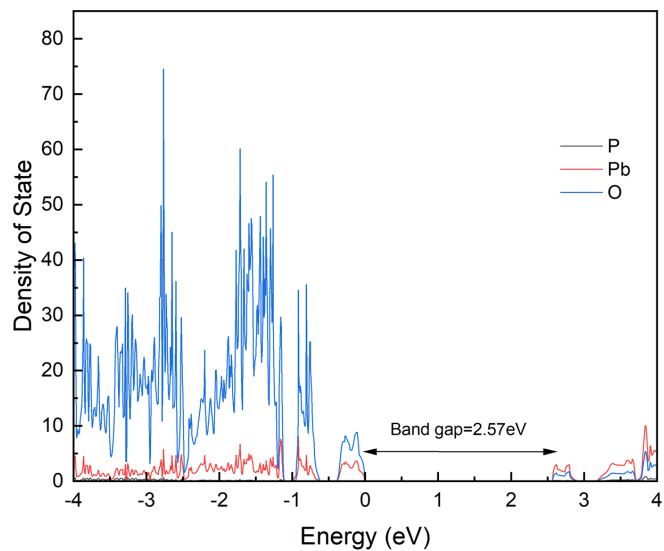

Figure 4 The electronic DOS for each element in the $Pb_{10}(PO_4)_6O$ compound

### 3.3.2 Electronic properties of $Pb_9Cu(PO_4)_6O$

After finishing the analysis of the electronic property of pure $Pb_{10}(PO_4)_6O$, we examined the electronic properties of one Cu atom substitution at both the Pb(1) and Pb(2) sites, leading to the formation of $Pb_9Cu(PO_4)_6O$:<Pb(1)> and $Pb_9Cu(PO_4)_6O$:<Pb(2)> (depicted in Figure 5 (a) and (b)). Furthermore, we investigated how the substitution of Cu atom will affect the electronic property of $Pb_9Cu(PO_4)_6O$ (Figure 6 and *Figure 7*).

In the case of $Pb_9Cu(PO_4)_6O$:<Pb(1)>, it is found that, after the doping of Cu, an isolated set of flat bands crosses the Fermi level with a maximum bandwidth of 0.13 eV. Such a narrow bandwidth is particularly indicative of strongly correlated bands. It has been reported that these notably flat bands observed align cohesively with the Cu-O coordination, revealing Cu-O bond lengths of 2.35 Å and 2.06 Å within the distorted trigonal prism. By way of comparison, in the context of cuprate superconductors, the Cu-O bond lengths typically range up to 2 Å for in-plane arrangements and up to 2.3 Å for apical configurations [10], giving further evidence of the unusual coordination and resulting band localization in this isolated Cu-*d* manifold. Figure 7 (a) reveals that the DOS with the energy range at the Fermi level is mainly contributed by the Cu-3d state, which closes the band gap and leads to an insulator-metal transition. However, the previously observed flat bands disappear near the Fermi level when the substitution site is switched to Pb(2), resulting in an opening of the band gap (around 1.3 eV) (Figure 7 (b)). Furthermore, the Cu substitution on Pb(2) sites will further lower the structure symmetry from $P6_3/m$ to P1 space group, which changes the high symmetry point of the first Brillouin zone [39]. The reopening of this gap reflects a substantial reduction in the electronic conductivity in $Pb_9Cu(PO_4)_6O$ compounds.

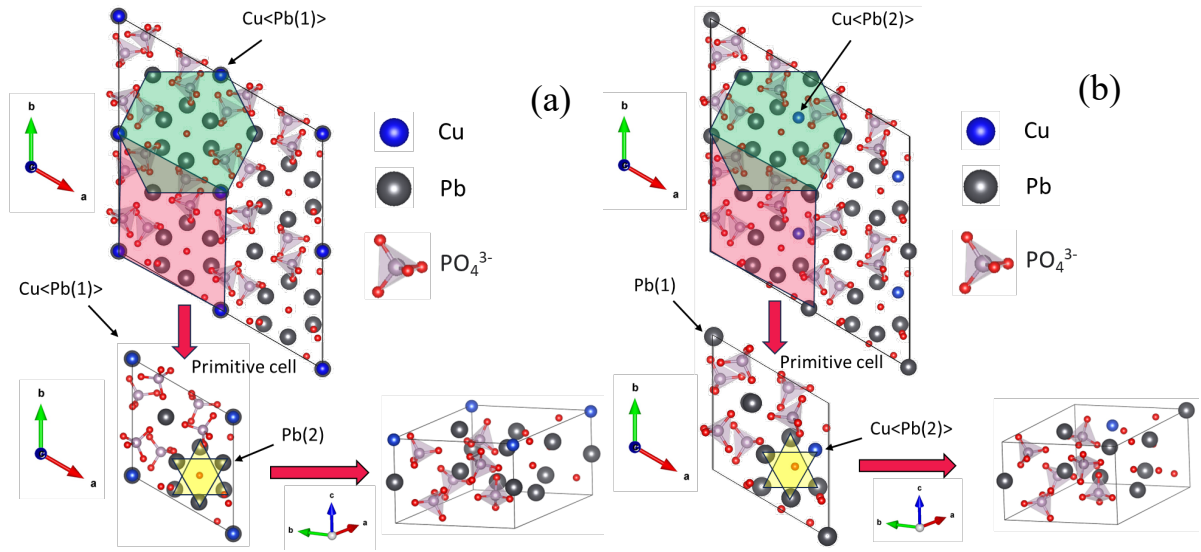

Figure 5 The structure configurations of $Pb_9Cu_1(PO_4)_6O$ compounds, where (a) $Pb_9Cu(PO_4)_6O:<Pb(1)>$, (b) $Pb_9Cu(PO_4)_6O:<Pb(2)>$

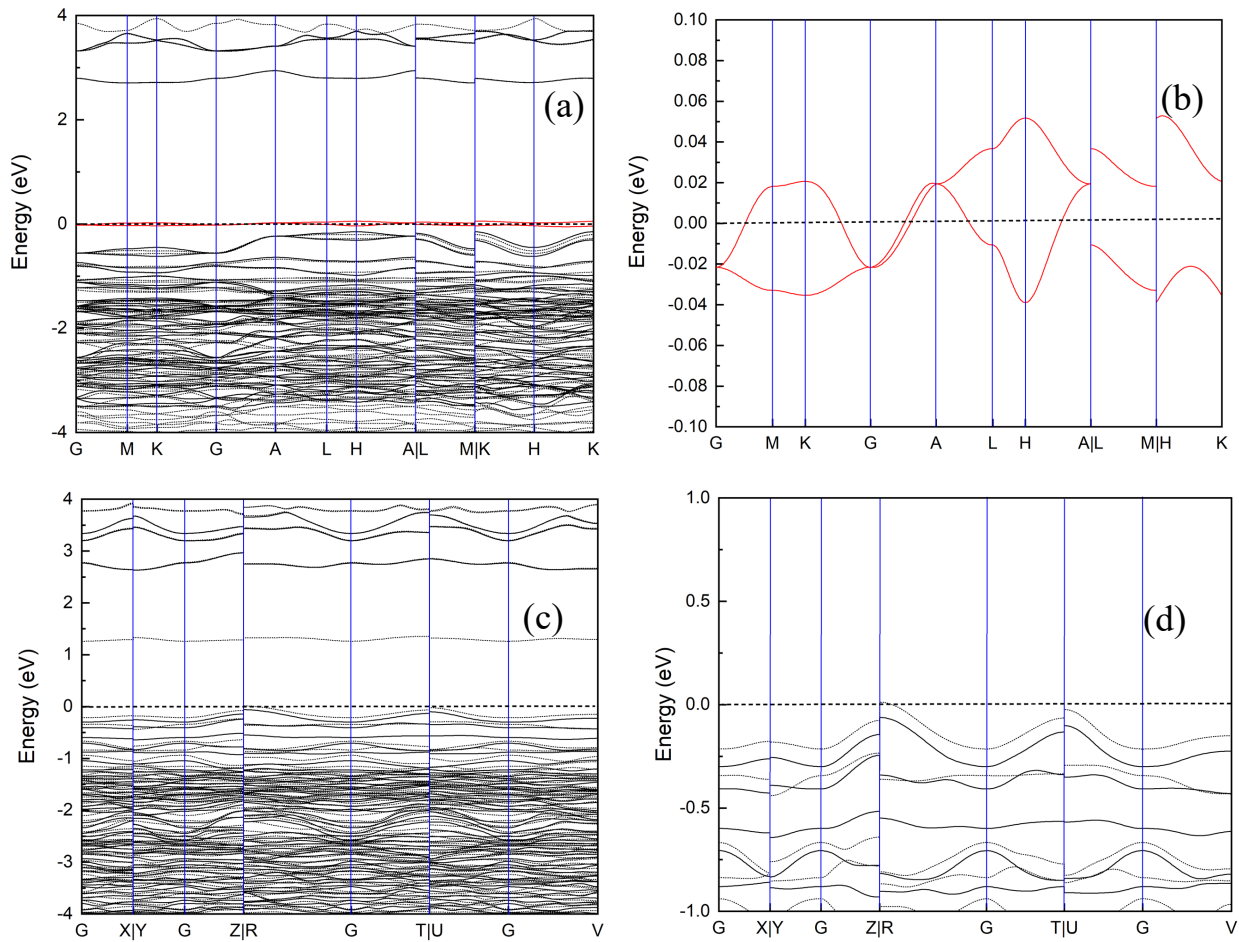

Figure 6 The Band structure of $Pb_9Cu_1(PO_4)_6O$ compounds, where (a) (b) $Pb_9Cu(PO_4)_6O:<Pb(1)>$, (c) (d) $Pb_9Cu(PO_4)_6O:<Pb(2)>$

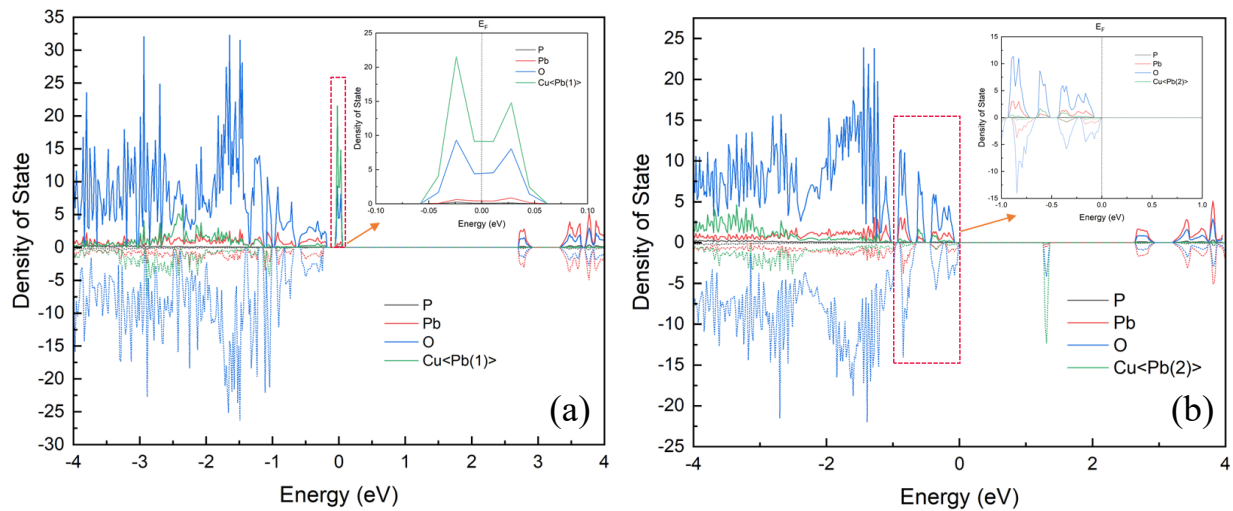

*Figure 7 The electronic DOS for each element in $Pb_9Cu_1(PO_4)_6O$, where (a) $Pb_9Cu(PO_4)_6O$:<Pb(1)>, (b) $Pb_9Cu(PO_4)_6O$:<Pb(2)>*

### 3.3.3 Electronic properties of $Pb_8Cu_2(PO_4)_6$ (x=2)

To gain insights into whether an elevated doping concentration of Cu would similarly give rise to this flat band, we also explored the case involving $Pb_8Cu_2(PO_4)_6$. It has been reported that the Cu dopant increases the possibility of the formation of oxygen vacancies [39]. Here, we considered removing the extra oxygen in the original $Pb_{10}(PO_4)_6O$ structure and then substituting two Pb atoms with two monovalent Cu atoms to balance the charge. Specifically, we examined the electronic properties of two Cu atoms substitution at both the Pb(1) sites, leading to the formation of $Pb_8Cu_2(PO_4)_6$:2Cu<2Pb(1)> Additionally, we also explored the scenario where two Cu atoms substituted one Pb(1) site and one Pb(2) site, respectively, resulting in the formation of $Pb_8Cu_2(PO_4)_6$:2Cu<1Pb(1), 1Pb(2)>. The fully relaxed structures are depicted in Figure 8. Similarly, the Cu substitution on Pb(2) site will further lower the structure symmetry from $P6_3/m$ to P1 space group, which shifts the high symmetry point of the first Brillouin zone [39].

The calculated band structures and electronic DOS of these two compounds are shown in Figure 9 and Figure 10. Overall, it is found that both Pb$_8$Cu$_2$(PO$_4$)$_6$:2Cu<2Pb(1)> and Pb$_8$Cu$_2$(PO$_4$)$_6$:2Cu<1Pb(1), 1Pb(2)> structures have bandgaps of 1.59 eV and 1.20 eV, respectively. Based on the DOS results, the flat valence bands near the Fermi level of these two compounds are dominated by the Cu-O bonding state. Compared with the Pb$_8$Cu$_2$(PO$_4$)$_6$:2Cu<2Pb(1)>, Pb$_8$Cu$_2$(PO$_4$)$_6$:2Cu<1Pb(1), 1Pb(2)> has fewer valence bands near the Fermi level.

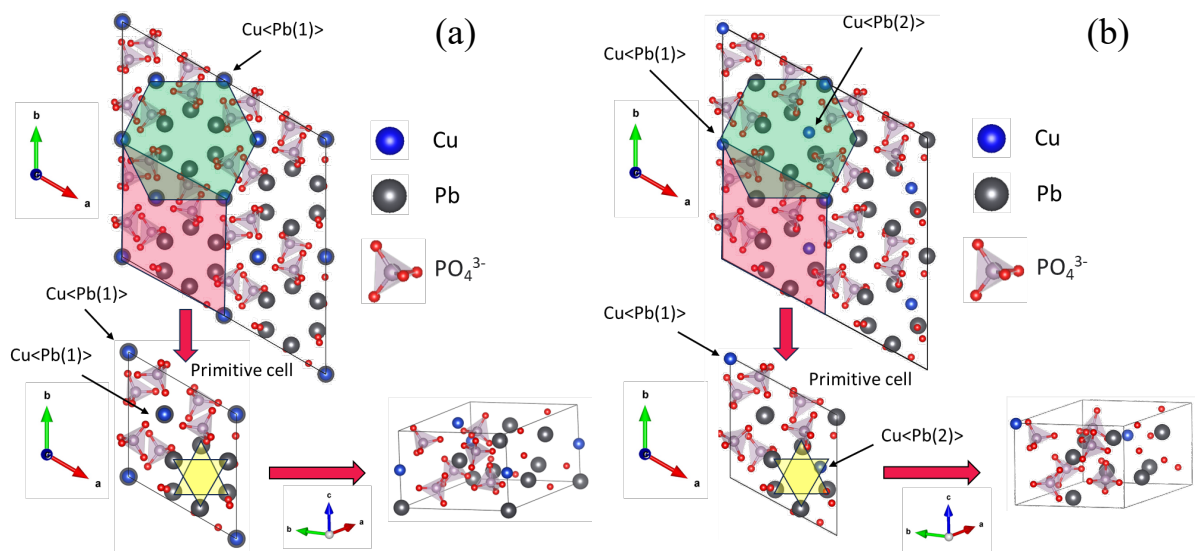

*Figure 8 The structure configurations of Pb$_9$Cu$_1$(PO$_4$)$_6$O compounds, where (a) Pb$_8$Cu$_2$(PO$_4$)$_6$:2Cu<2Pb(1)>, (b) Pb$_8$Cu$_2$(PO$_4$)$_6$:2Cu<1Pb(1), 1Pb(2)>.*

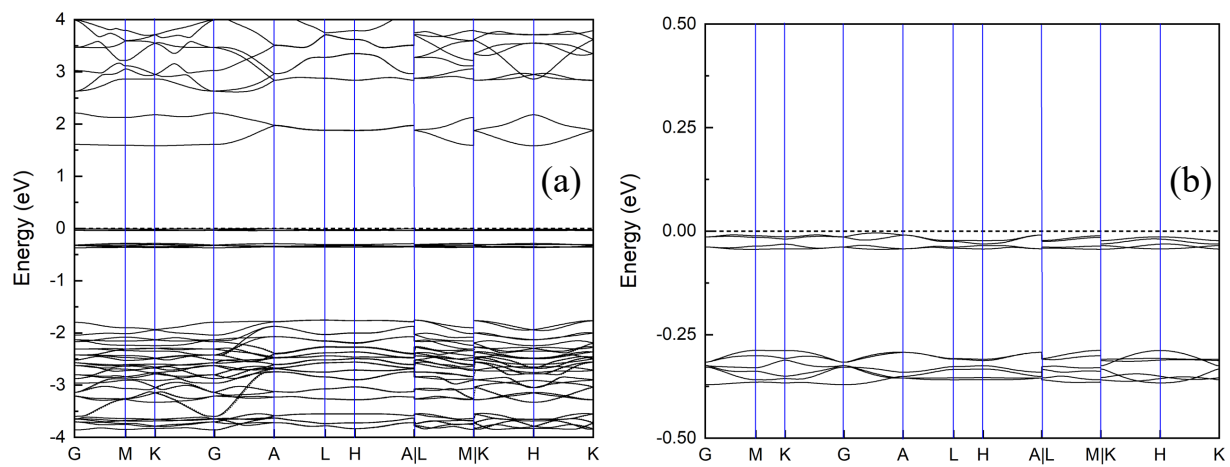

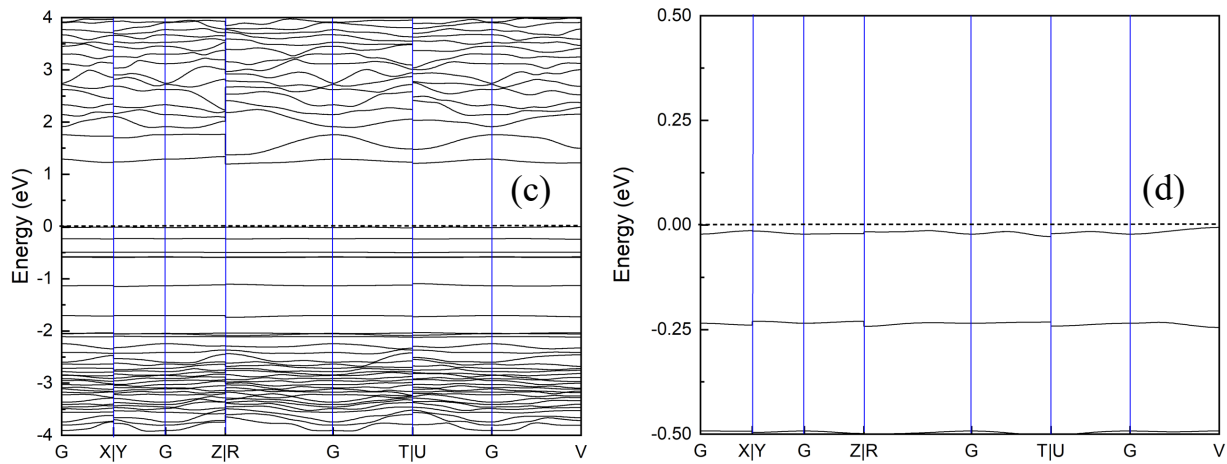

Figure 9 The Band structure of $Pb_8Cu_2(PO_4)_6$ (x=2) compounds, where (a) (b) $Pb_8Cu_2(PO_4)_6$:2Cu<2Pb(1)>, (c) (d) $Pb_8Cu_2(PO_4)_6$:2Cu<1Pb(1), 1Pb(2)>

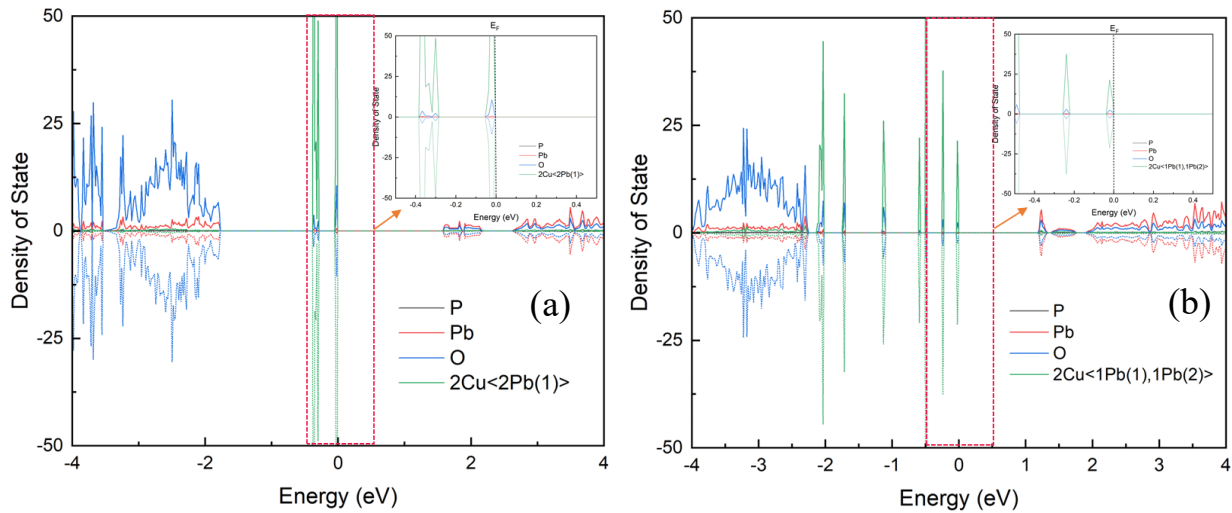

Figure 10 The electronic DOS for each element in $Pb_8Cu_2(PO_4)_6$ (x=2), where (a) $Pb_8Cu_2(PO_4)_6$:2Cu<2Pb(1)>, (b) $Pb_8Cu_2(PO_4)_6$:2Cu<1Pb(1), 1Pb(2)>.

3.4. Electronic conductivity of $Pb_{10-x}Cu_x(PO_4)_6O$ (x=0,1) and $Pb_8Cu_2(PO_4)_6$ (x=2)

The electrical conductivity of $Pb_{10-x}Cu_x(PO_4)_6O$ can be estimated based on the Boltzmann transport theory, which is written as:

$$\sigma = \frac{e^2}{Vk_BT}\int_{-\infty}^{\infty} f(1-f)\Xi(\varepsilon)d(\varepsilon) \tag{2}$$

https://arxiv.org/abs/2308.13938

where $e$ is the elementary charge, $V$ is the volume, $T$ represents the temperature, $\varepsilon$ is the volume one-electron energy, $f$ denotes the Fermi distribution, and $\Xi(\varepsilon)$ is defined as the transport function [45, 46]. $\Xi(\varepsilon)$ is a tensor with components [47]:

$$\Xi^{\alpha\beta}(\varepsilon) = \int \sum_i \tau_{i,k} v_i^\alpha(k) v_i^\beta(k) \delta[\varepsilon - \varepsilon_i(k)] \frac{dk}{8\pi^3} \quad (3)$$

where $\alpha$ and $\beta$ are the indices labeling the Cartesian axis, $i$ is the one-electron band index, $\tau_{i,k}$ denotes electron relaxation time, and the electron group velocity $v_i^\alpha$ represents the gradient of electron band energy with respect to $k$, namely:

$$v_i^\alpha(k) = \frac{\partial \varepsilon_i(k)}{\hbar k^\alpha} \quad (4)$$

The relaxation time $\tau_{i,k}$ is estimated based on the model proposed by Wang et al. [47]:

$$\tau_{i,k} = x \frac{\hbar}{2T} \sqrt{\frac{n}{k_B c_{el}}} \quad (5)$$

$$c_{el} = \frac{1}{k_B T^2} \int_{-\infty}^{\infty} (\Delta\varepsilon)^2 (1-f) f D(\varepsilon) d\varepsilon \quad (6)$$

$$D(\varepsilon) = \int \sum_i \delta[\varepsilon - \varepsilon_i(k)] \frac{dk}{8\pi^3} \quad (7)$$

$$n = \int_{-\infty}^{\infty} (1-f) f D(\varepsilon) d\varepsilon \quad (8)$$

where $x$ is treated as a material constant, and the current work use $x=0.5$ for the case of $Pb_{10-x}Cu_x(PO_4)_6O$. $\Delta\varepsilon$ represents the energy uncertainty. $D(\varepsilon)$ denotes the electronic DOS.

The calculated electronic conductivities and resistivities of these five structures are shown in *Figure 11* and *Figure 12*. The pure $Pb_{10}(PO_4)_6O$ (plotted as solid black lines) has been selected as the reference sample for the purpose of comparison. The electronic conductivity of all structures

https://arxiv.org/abs/2308.13938

tends to decrease as the temperature increases due to a concurrent increase in electronic resistivity. The decreasing slope of the electronic conductivity becomes much smoother as the increase of temperature. With the exception of $Pb_9Cu(PO_4)_6O:<Pb(1)>$, all the other compounds demonstrate limited electronic conductivities due to the presence of a band gap. In contrast, $Pb_9Cu(PO_4)_6O:<Pb(1)>$ exhibits comparatively higher electronic conductivity attributed to an isolated set of flat bands that intersect the Fermi level. For the case of $Pb_8Cu_2(PO_4)_6$, it is observed that, compared with the pure $Pb_{10}(PO_4)_6O$, the increasing of the Cu substitution will exponentially increase the electronic resistivity and lower the conductivity, indicating that the formation of $Pb_8Cu_2(PO_4)_6$ is detrimental to the electronic properties of the material.

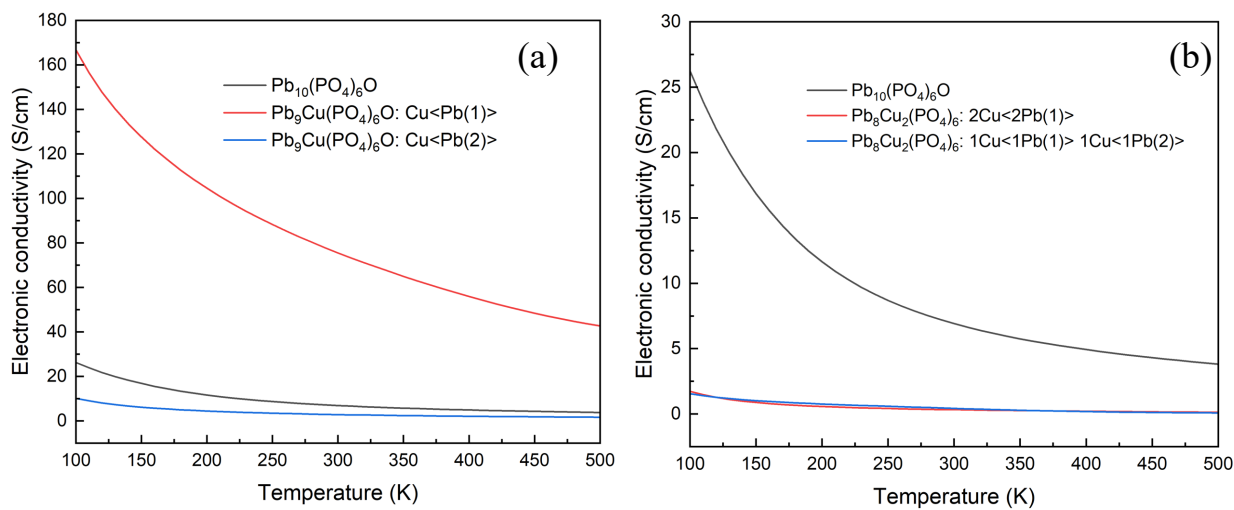

Figure 11 The electronic conductivity curve of (a) $Pb_{10-x}Cu_x(PO_4)_6O$ (x=0,1) and (b) $Pb_8Cu_2(PO_4)_6$ (x=2)

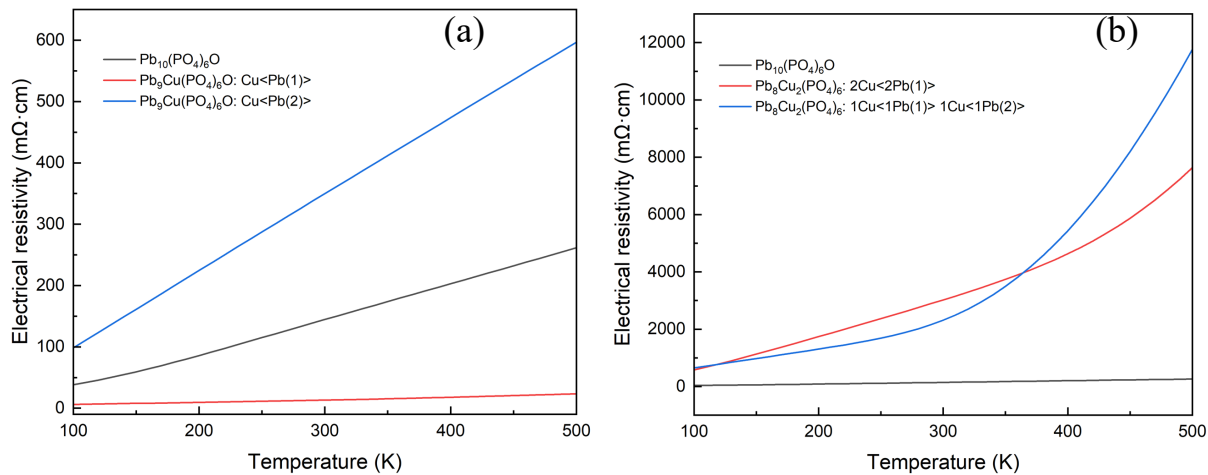

*Figure 12 The electronic resistivity curve of (a) $Pb_{10-x}Cu_x(PO_4)_6O$ (x=0,1) and (b) $Pb_8Cu_2(PO_4)_6$ (x=2)*

## 4. Discussions on LK-99

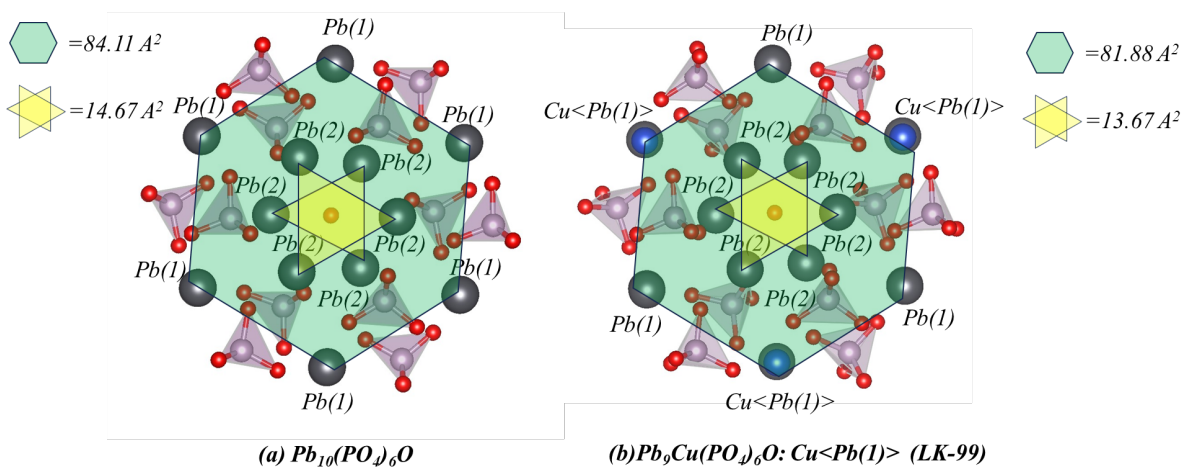

*Figure 13 The atomic distance analysis of pure $Pb_{10}(PO_4)_6O$ and $Pb_9Cu(PO_4)_6O:<Pb(1)>$ (LK-99)*

After systematically investigating the structural and electrical properties of $Pb_{10-x}Cu_x(PO_4)_6O$ (x=0,1) and $Pb_8Cu_2(PO_4)_6$, it is suggested that $Pb_9Cu(PO_4)_6O:Cu<Pb(1)>$ (LK-99) display a flat isolated band [17]. Meanwhile, our calculations as shown in *Figure 13* indicates that, after the doping of Cu, the area of hexagonal frame constructed by Pb(1) atoms have undergone a reduction from 84.11 Å$^2$ to 81.88 Å$^2$, reflecting a contraction of approximately 2.65%. Similarly, the area of two mutually inverted triangles formed by Pb(2) atoms have diminished from 14.67 Å$^2$ to 13.67 Å$^2$, indicating an approximate shrinkage of 6.8%. This trend shows good agreement with the reported formation of the superconducting quantum wells (SQWs) by the distortion through the lattice contraction for LK-99 [15, 16].

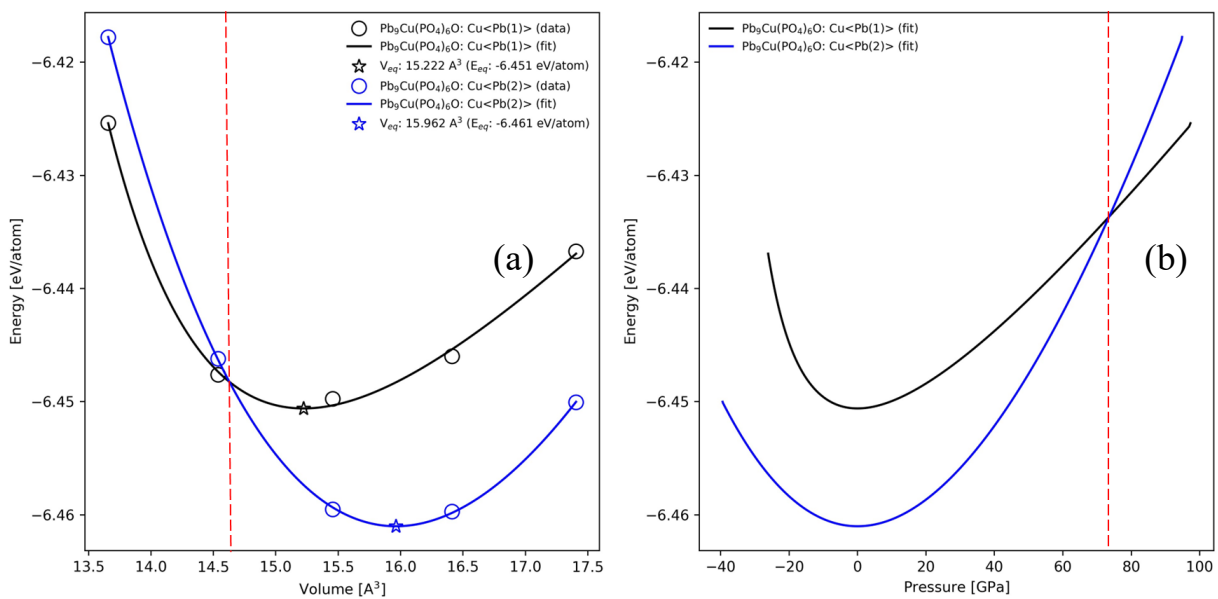

*Figure 14 The equation of state diagram of Pb$_9$Cu(PO$_4$)$_6$O with two different structures (a) E-V diagram (b) E-P diagram*

However, based on the total energy calculations, it is observed that the Cu substitution on Pb(2) sites is more energetically favorable than Pb(1) sites by around 0.010 eV per atom (0.9645 kJ/mol). This result suggests a challenge in the synthesis process when attempting to substitute Cu into the suitable site to achieve a bulk superconducting sample [17] unless carried out under extremely high pressures. Here, we plotted the equation of state diagram of both cases in *Figure 14*. It is demonstrated that the energy difference between these two structures will significantly decrease with the rising pressure, and LK-99 will be more stable when the pressure is larger than 73 GPa.

## 5. Conclusions

In this comprehensive investigation, we systematically explored the stability and electronic properties of $Pb_{10-x}Cu_x(PO_4)_6O$ (x=0,1) and $Pb_8Cu_2(PO_4)_6$ (x=2), employing the *ab initio* approach. The key findings can be summarized as follows:

(1). The electronic property result shows a set of isolated flat bands crossing the Fermi level solely emerges in $Pb_9Cu(PO_4)_6O$: Cu<Pb(1)>, where Pb(1) denotes the substitution site, which is corresponding to LK-99. Such flat bands behavior hints $Pb_9Cu(PO_4)_6O$: Cu<Pb(1)> can be a superconductor candidate, even though it may not be a room temperature superconductor. Meanwhile, the remaining four parent compounds exhibit insulating behavior characterized by wide band gaps, which explains why very different properties were claimed from the experimental observation from different groups. In practice, a strict control of sample stoichiometry and crystal structure would be the key.

(2). The EOS calculations reveal a noteworthy lattice contraction within LK-99, leading to an approximate volumetric reduction of 5.5% at 0K. Especially the lattice contraction around 6.8% in the area of two mutually inverted triangles formed by Pb(2) atoms may facilitate the formation of cooper pairs.

(3). The XRD spectra generated from the DFT simulations confirmed the presence of LK-99, claimed from the experimental XRD results. Meanwhile, the presence of an additional experimental peak between peaks d and e, as observed in the original publication [16], implies the existence of secondary phases. One plausible interpretation suggests the potential occurrence of the $Pb_8Cu_2(PO_4)_6:2Cu<2Pb(1)>$ phase.

(4). With the DFT simulations at 0K, we confirmed that substituting one Cu atom on Pb(2) sites ($Pb_9Cu(PO_4)_6O:Cu<Pb(2)>$) holds a slight energy advantage over Pb(1) sites ($Pb_9Cu(PO_4)_6O:Cu<Pb(1)>$), by approximately 0.010 eV per atom (equivalent to 0.9645 kJ/mol). It is possible that LK-99 is a high temperature stable phase and will undergo a phase transformation with the drop of temperature. Intriguingly, LK-99 with Cu substitution on Pb(1) sites becomes stable under pressures exceeding 73 GPa, i.e., the LK-99 phase transformation temperature can be tuned by increasing the synthesis pressure.

(5). The electronic conductivity for LK-99 calculated based on the Boltzmann transport theory underscores significantly enhanced conductivity in comparison to the other four parent compounds. However, it falls short of achieving comparability with the conductivity levels observed in metals or other advanced oxide conductors.

## 6. Acknowledgment

The work was supported by the Department of Energy under Award Number: DE-FE0032116 and Advanced Cyberinfrastructure Coordination Ecosystem: Services & Support (ACCESS) program Award Numbers TG-DMR190004.

https://arxiv.org/abs/2308.13938